\documentclass{PoS}
\usepackage{bm}

\title{Dark matter/new physics searches at BESIII}

\ShortTitle{Dark matter/new physics searches at BESII}

\author{\speaker{Vindhyawasini Prasad}\thanks{Work supported in part by the National Natural Science Foundation of China (NFSC) under contract No. 11705192, and   $64^{th}$ batch of Postdoctoral Science Fund Foundation under contract No. 2018M642516.}\\
        Department of Modern Physics, University of Science $\&$ Technology of China, Hefei 230026  State Key Laboratory of Particle Detection and Electronic, Beijing 100049, Hefei 230026, China\\
        E-mail: \email{vindy@ustc.edu.cn}}

\abstract{Many new physics models beyond the Standard Model, motivated by the recent astrophysical anomalies, include the possibility of several new types of light weak-interacting particles. Typical models, such as Next-to-Minimal Supersymmetric Standard Model and  Hidden Dark-sector Model,  predict the light Higgs and dark bosons, respectively.  The masses of these particles are expected to be a few GeV and thus making them accessible at BESIII. BESIII has recently explored the possibility of light Higgs and dark bosons in several decay modes using the data-sets collected at $J/\psi$, $\psi(3686)$ and $\psi(3770)$ resonances. The $J/\psi$ data have further utilized to search for invisible decays of light vector ($V=\omega, \phi$) and pseudo-scalar ($P=\eta, \eta'$) mesons via $J/\psi \to VP$ decays. This report summarizes the recent results of the BESIII experiment on these new physics topics.}

\FullConference{ALPS 2019 An Alpine LHC Physics Summit \\
		April 22 - 27, 2019\\
		Obergurg, Austria}

\begin{document}

\section{Introduction} Many astrophysical observations strongly suggest the existence of dark matter~\cite{anomaly}, which nature is still mysterious in modern physics. Dark matter neither emits nor absorbs the electromagnetic radiation and  its presence can only be inferred via gravity.  New physics models beyond the Standard Model (SM) motivate a new type of {\lq hidden dark-sector\rq} under which the WIMP like dark matter particles are charged via a new type of force carrier~\cite{hidden}. The corresponding gauge field can couple to the ordinary matter via {\lq\lq portals\rq\rq}~\cite{essig}, which could be scalar, pseudoscalar, vector  or spin-1/2 fermions. The masses of these new particles are expected to be a few GeV to satisfy the constraints of recent experimental anomalies~\cite{anomaly, amu},  and thus making them accessible via  high intensity electron-positron collider experiments, such as BESIII experiment.  BESIII,  an $e^+e^-$ collider experiment running at tau-charm region, has collected a huge amount of data at several energy points between 2.0-4.6 GeV, including $J/\psi$, $\psi(3686)$ and $\psi(3770)$ resonances, to study the light hadron spectroscopy and search for new physics beyond the SM. This report summarizes the recent results of the BESIII experiment related to the new physics/dark matter searches. 

\section{Search for dark  photon}
The dark photon ($\gamma'$) is a new type of force carrier in the simplest scenario of an Abelian $U(1)$ gauge field.  It  couples to the SM photon via kinetic mixing strength,  defined as $\epsilon^2=\alpha'/\alpha$, where $\alpha'$ and $\alpha$ are the fine structure constants in the dark and SM sectors, respectively~\cite{hidden}. The mass of the $\gamma'$ is expected to be a few GeV to satisfy the constraints of recent astrophysical anomalies~\cite{anomaly}, as well as the observed deviation in the muon anomalous magnetic moment up to the level of $(3-4)\sigma$ between theory and experiment~\cite{amu}.   A series of experiments have performed the searches for $\gamma'$ and reported only null results so far  with an exclusion limit on $\epsilon$  to be below $10^{-3}$~\cite{exclusion, benedikt}.

\begin{figure}[!htp]
\begin{center}
\includegraphics[width=0.48\textwidth]{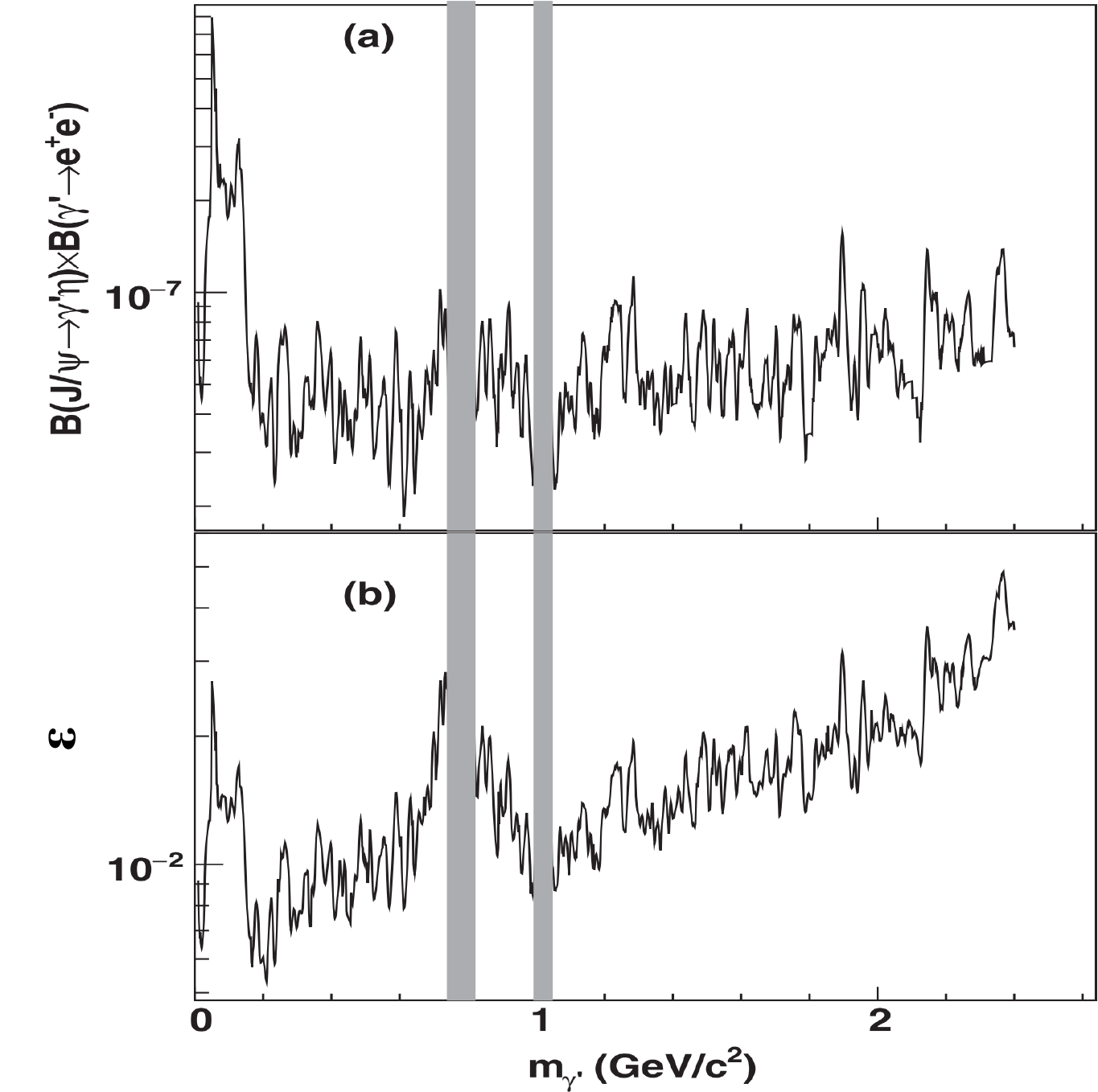}
\includegraphics[width=0.48\textwidth]{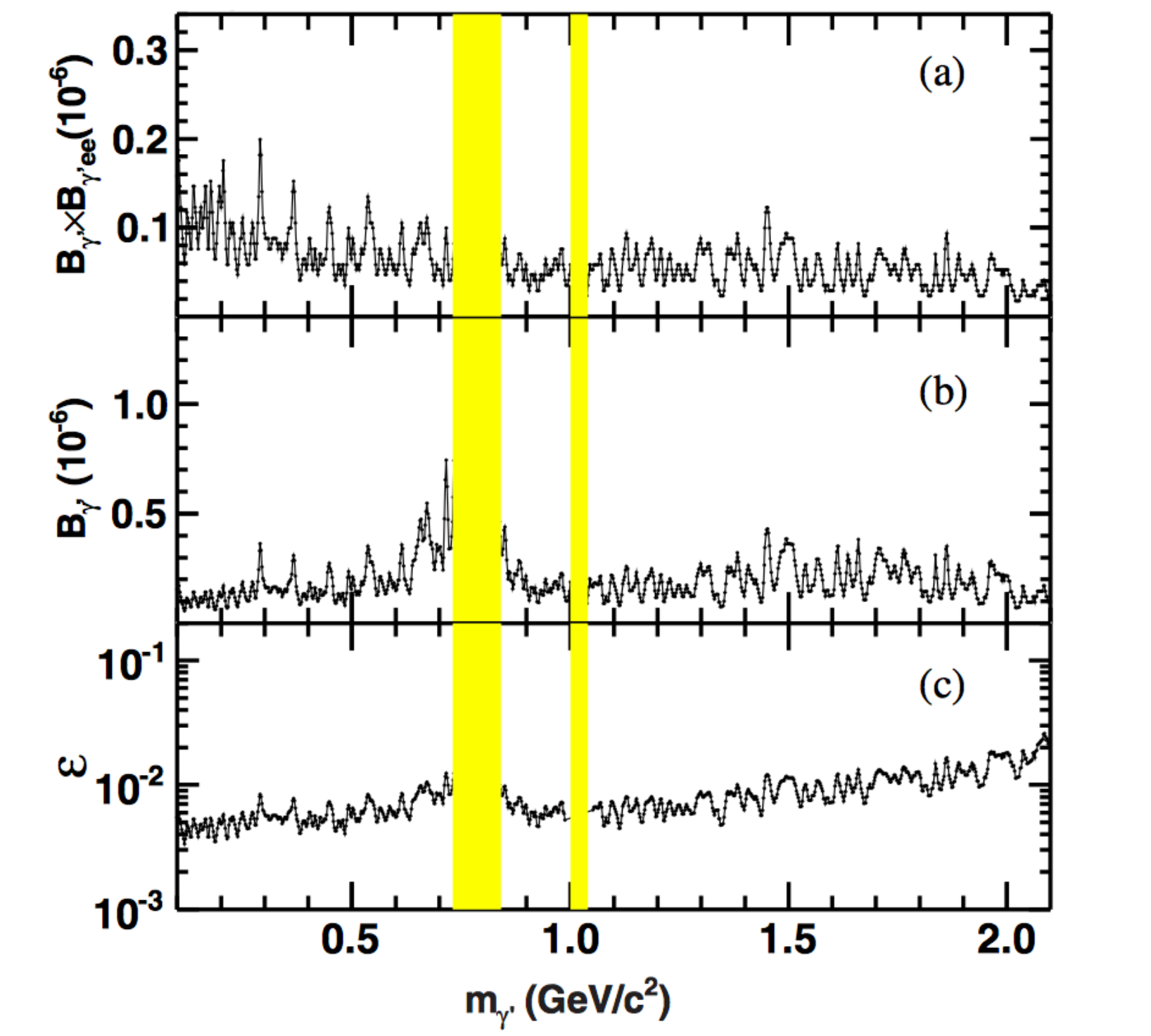}
\caption{ The $90\%$ C.L. upper limits on the branching fractions $\mathcal{B}(J/\psi \to \gamma' \eta(')) \times \mathcal{B}(\gamma' \to e^+e^-)$ (top) and the kinetic mixing strength $\epsilon$ (bottom). The left plots are for $J/\psi \to \gamma' \eta$ decay and right plots for $J/\psi \to \gamma' \eta'$ decay. The right hand side figure also shows the plot of $90\%$ C.L. upper limit on $\mathcal{B}(J/\psi \to \gamma' \eta')$.}
\label{ULBF}
\end{center}
\end{figure}

BESIII has recently searched for di-electron decays of a $\gamma'$ through $J/\psi \to \gamma' \eta (')$ using 1.3 billion of $J/\psi$ events, where $\eta$ ($\eta'$) is reconstructed from $\eta \to \pi^+\pi^-\pi^0$ , $\pi^0 \to \gamma \gamma$ and $\eta \to \gamma \gamma$ ($\eta' \to \gamma \pi^+\pi^-$ and $\eta' \to \eta \pi^+\pi^-$, $\eta \to \gamma \gamma$)~\cite{vindy}. The search for a narrow $\gamma'$ resonance is performed in the $m_{e^+e^-}$ distribution of the data. But, we exclude the $\omega$ and $\phi$ mass regions in the $m_{e^+e^-}$ spectrum  from the $\gamma'$ searches due to their resolutions are compatible with the $\gamma'$ mass resolution. No evidence of $\gamma'$ production is found.  The $90\%$ confidence level upper limits on product branching fractions $\mathcal{B}(J/\psi \to \eta (') \gamma') \times \mathcal{B}(\gamma' \to e^+e^-)$ and  $\epsilon$ are set to be up to the level of $10^{-8}$ and $10^{-3}$, respectively, depending upon the $\gamma'$ mass points (Figure~\ref{ULBF}).

\begin{figure}[!htp]
\centering
\includegraphics[width=0.48 \textwidth]{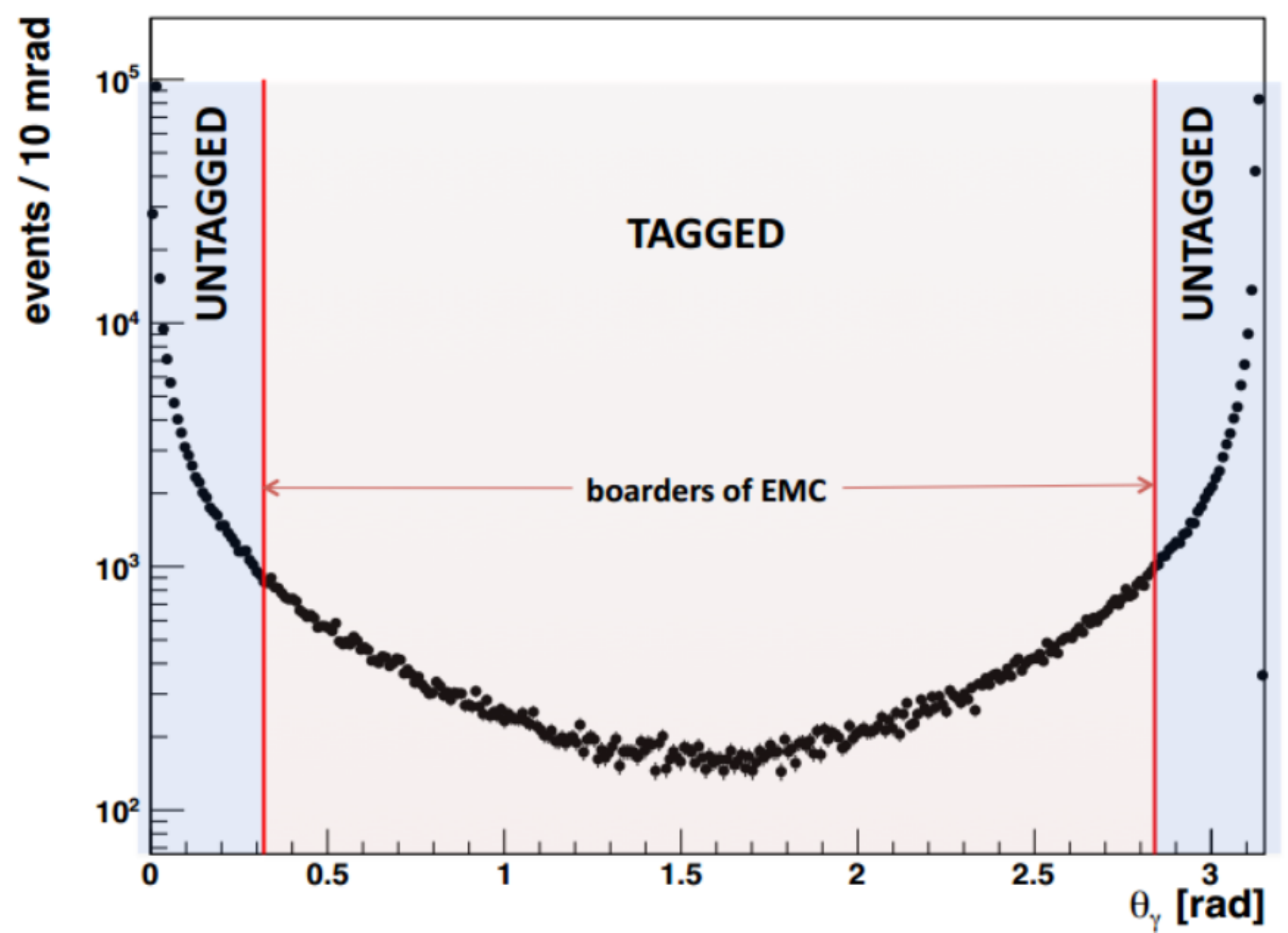}
\includegraphics[width=0.48  \textwidth]{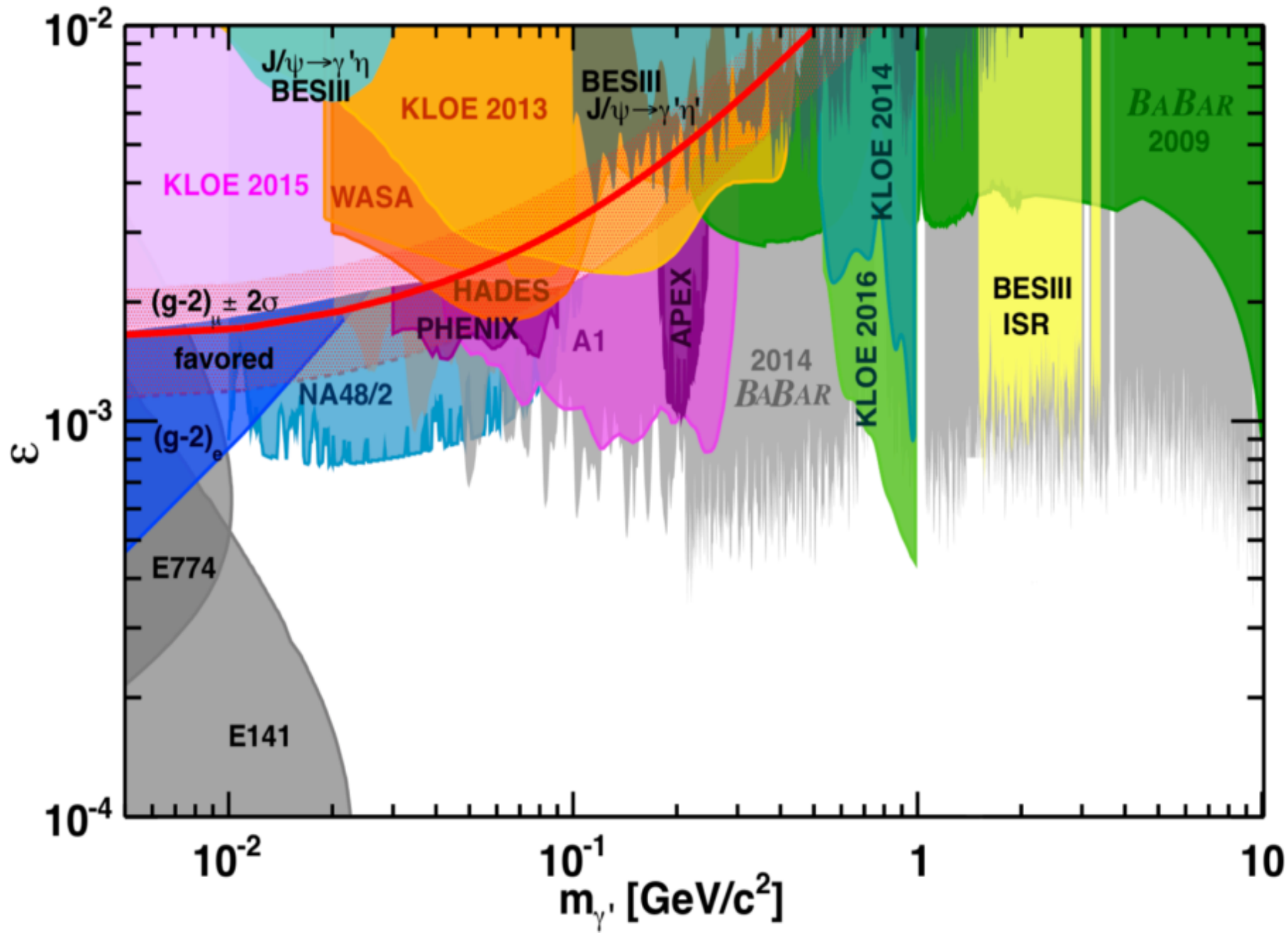}

\caption{(Left) Polar angle ($\theta_{\gamma}$) distribution of the ISR photon and (right) the $90\%$ C.L. upper limits on $\epsilon$ as a function of $m_{\gamma'}$ for various experimental measurements~\cite{exclusion} including the BESIII~\cite{benedikt, vindy}. BESIII detector acceptance region for $\theta_{\gamma}$ is [0.376, 2.765] radian. In untagged photon method, the ISR photon is required to be not detected in the EMC detection region. }
\label{epsilon}
\end{figure}

BESIII has also performed the search for dark photon via initial-state-radiation (ISR) process $e^+e^- \to \gamma_{ISR} \gamma'$, $\gamma' \to l^+l^-$ ($l=e,\mu$)  using 2.93 fb$^{-1}$ $\psi(3770)$ data~\cite{benedikt}. This search uses an untagged photon method in which ISR photon is not detected in the BESIII electromagnetic calorimeter acceptance region (Figure~\ref{epsilon} (left)). The background in this search is mainly dominated by the ISR processes of $e^+e^- \to \gamma l^+l^-$ $(l=e,\mu)$.  No any obvious enhancement above these backgrounds is seen in di-lepton invariant mass spectrum, and a $90\%$ C.L. upper limit on $\epsilon$ is set in the mass range between 1.5 and 3.4 GeV/$c^2$. The obtained upper limits in tested mass range are compatible with the BaBar measurement (Figure~\ref{epsilon} (right)). 

\section{invisible decays of light mesons}
Neutrinos ($\nu$) are the only invisible particles within the SM that don't interact with the particle physics detector. Only the four momenta of missing particles in a  decay mode is used to infer the presence of this particle. Quarknonium, which is a composition of a quark ($q$) and its own anti-quark ($\overline{q}$), annihilates into a neutrino-pair rarely via virtual $Z^0$ boson. The branching fractions of such a rare decays might enhance in the presence of light dark matter (LDM) particles. Ref.~\cite{McElrath} predicts the branching fractions of the invisible $\chi$ decays of various quarkonium states  while assuming the same cross-section for the time reversed processes $\sigma(q\overline{q} \to \chi\chi)  \simeq \sigma(\chi \chi \to q\overline{q})$. BaBar~\cite{babar} and BESII~\cite{bes2} experiments have set one of most stringent upper limits on the invisible decays of $\Upsilon(1S)$ and $J/\psi$ mesons, respectively, which are still above SM predictions. By using 225 million $J/\psi$ events, BESIII has analyzed $J/\psi \to \phi \eta(')$ decays and determined upper limits at $90\%$ C.L. to be $2.6 \times 10^{-4}$ for the ratio $\frac{\mathcal{B}(\eta \to \rm{invisible})}{\mathcal{B}(\eta \to \gamma \gamma)}$ and $2.4 \times 10^{-2}$ for $\frac{\mathcal{B}(\eta' \to {\rm invisible})}{\mathcal{B}(\eta' \to \gamma \gamma)}$~\cite{etatoinv}.

\begin{figure}[!htp]
\centering
\includegraphics[width=0.32 \textwidth]{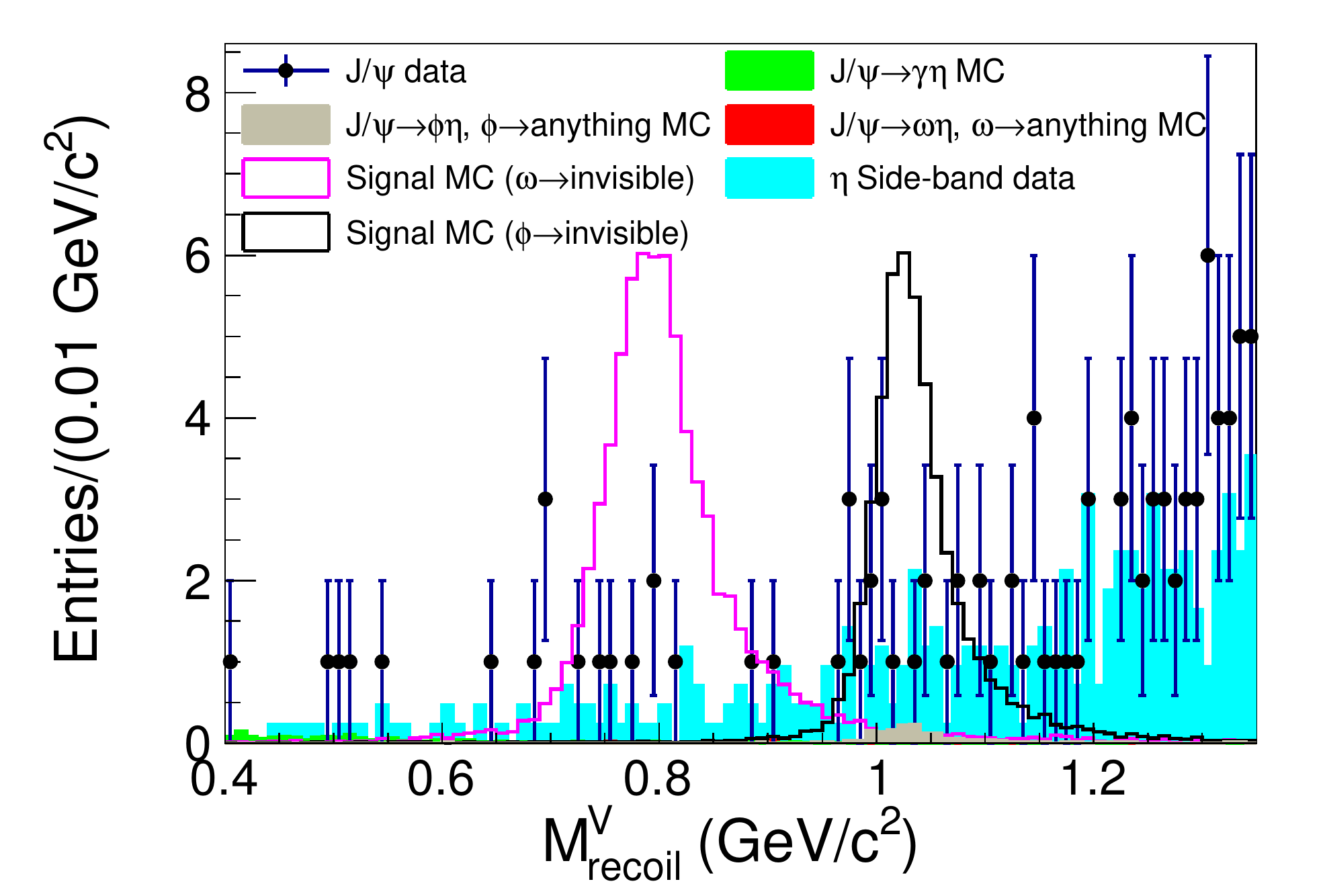}
\includegraphics[width=0.65 \textwidth]{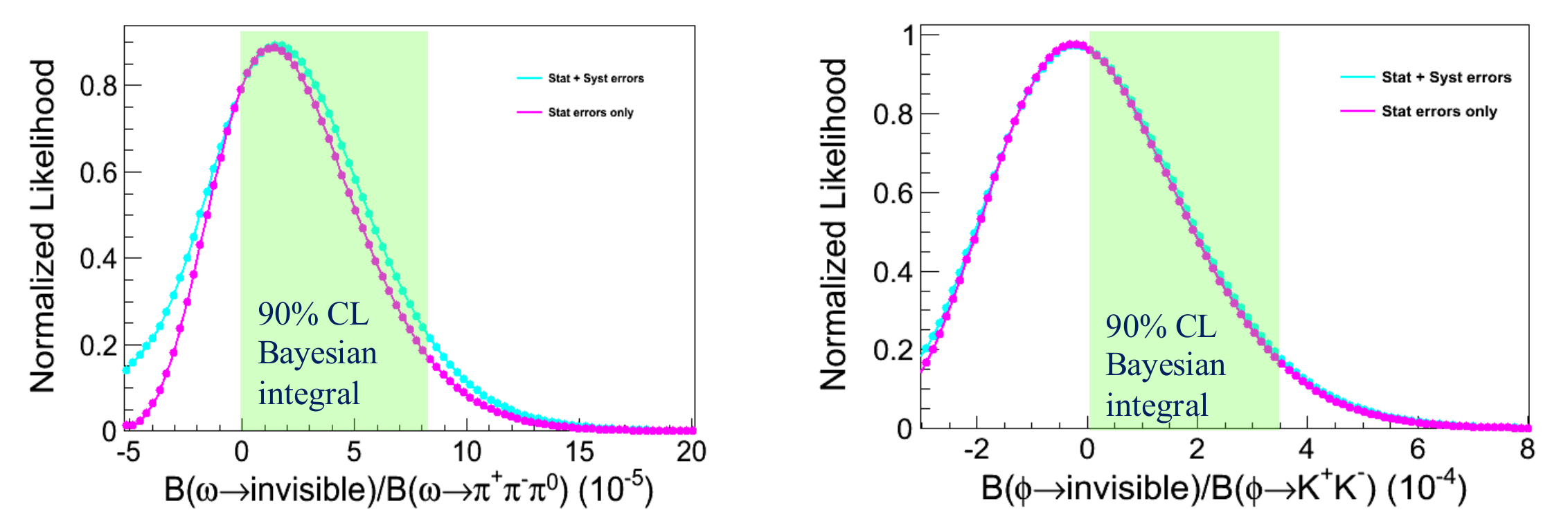}

\caption{The $M_{\rm recoil}^V$  distribution for data (black dots with error bars), signal MC samples (pink and black histograms for $\omega$ and $\phi$  decays, respectively) and various expected backgrounds shown by various colored histograms (left),  and  the normalized likelihood versus $\frac{\mathcal{B}(\omega \to invisible)}{\mathcal{B}(\omega \to \pi^+\pi^-\pi^0)}$ (middle) and $\frac{\mathcal{B}(\phi \to invisible)}{\mathcal{B}(\phi \to K^+K^-)}$ (right) including the statistical uncertainty only (pink) and  both systematic and statistical uncertainties (cyan). }
\label{mrec}
\end{figure}

The SM predicts $\mathcal{B}( \omega \to \nu \overline{\nu})=(2.79 \pm 0.05) \times 10^{-13}$ and $\mathcal{B}( \phi \to \nu \overline{\nu})=(1.67 \pm 0.02) \times 10^{-11}$~\cite{dngao}, but the presence of LDM may enhance the branching fractions $\omega/\phi \to \chi \chi$ up to the level of $10^{-8}$~\cite{McElrath}.  BESIII has recently  performed the search for invisible decays of $\omega$ and $\phi$ meson in $J/\psi \to V \eta$  decays using 1.3 billion $J/\psi$ events, where  $\eta$  is reconstructed by $\pi^+\pi^-\pi^0$ with $\pi^0 \to \gamma \gamma$ in the final state~\cite{invis}. The mass recoiling against $\eta$, $M_{\rm recoil}^V=\sqrt{(E_{\rm CM}-E_{\pi^+\pi^-\pi^0})^2-\overrightarrow{p}_{\pi^+\pi^-\pi^0}^2}$, is used to search for invisible decays of $\omega$ and $\phi$ mesons (Figure~\ref{mrec} (left)). No evidence of the significant invisible signal events is observed, and the upper limits on the ratio of branchings at the $90\%$ C.L. are measured to be $\frac{\mathcal{B}(\omega \to invisible)}{\mathcal{B}(\omega \to \pi^+\pi^-\pi^0)} < 8.0 \times 10^{-5}$ and  $\frac{\mathcal{B}(\phi \to invisible)}{\mathcal{B}(\phi \to K^+K^-)} < 3.4 \times 10^{-4}$ using the Bayesian approach~\cite{pdg} (Figure~\ref{mrec} (middle  and right)). The $90\%$ C.L. upper limits on $\mathcal{B}(\omega \to \rm invisible)$ and $\mathcal{B}(\phi \to \rm invisible)$ also set to be less than $7.2 \times 10^{-5}$ and $1.7 \times 10^{-4}$, respectively, for the first time using the world  average values of $\mathcal{B}(\omega \to \pi^+\pi^-\pi^0)$ and $\mathcal{B}(\phi \to K^+K^-)$~\cite{pdg}.

\section{Search for a \bm{$CP$}-odd like Higgs boson}
A light Higgs boson is predicted by many models beyond the SM including Next-to-Minimal Supersymmetric Model (NMSSM)~\cite{nmssm}. The Higgs sector of the NMSSM contains seven Higgs bosons, among them there is a $CP$-odd light Higgs boson ($A^0$) whose mass is expected to a few GeV. Such a low-mass Higgs boson can be accessible via $J/\psi \to \gamma A^0$ using the data of BESIII experiment collected at $J/\psi$ and $\psi(3686)$ resonances~\cite{wilczek}. Coupling of the Higgs field to the charm {\lq $c$\rq} (bottom {\lq $b$\rq}) quark-pair ($g_{b/c}$) is proportional to $\cos\theta_A/\tan\beta$ ($\cos\theta_A  \tan\beta$),  where $\tan\beta$ is a standard SUSY parameter and $\cos\theta_A(=|\sqrt{g_bg_c}|)$ is a mixing parameter between singlet and doublet components of the $A^0$~\cite{Gunion, domingo}. The branching fraction of $J/\psi \to \gamma A^0$ is predicted to be in the range of $10^{-9}-10^{-7}$, depending up on the $A^0$ mass, coupling and the NMSSM parameters~\cite{Gunion, domingo}. By using 106 million $\psi (3686)$ events, BESIII has previously set $90\%$ C.L. upper limits on product branching fractions $\mathcal{B}(J/\psi \to \gamma A^0)\times \mathcal{B}(A^0 \to \gamma \gamma)$ to be in the range of $(4 - 210) \times 10^{-7}$ depending upon the $A^0$ mass point for $m_{A^0} < 3.0$ GeV/$c^2$~\cite{A0psip}, where  a $J/\psi$ sample is selected by tagging a pion-pair from $\psi(3686) \to \pi^+\pi^- J/\psi$ transitions.  

\begin{figure}[!htp]
\centering
\includegraphics[width=0.50 \textwidth]{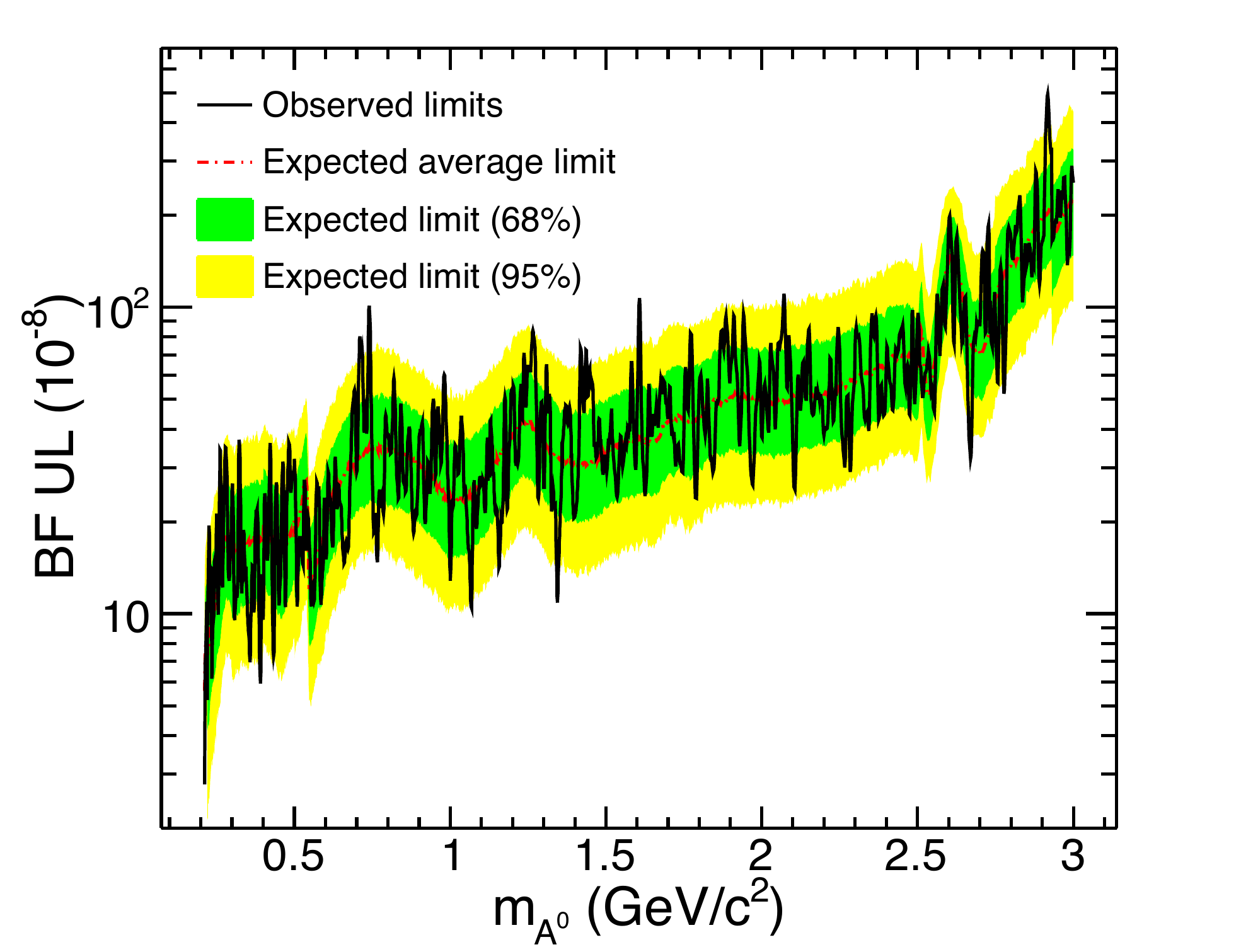}
\includegraphics[width=0.46 \textwidth]{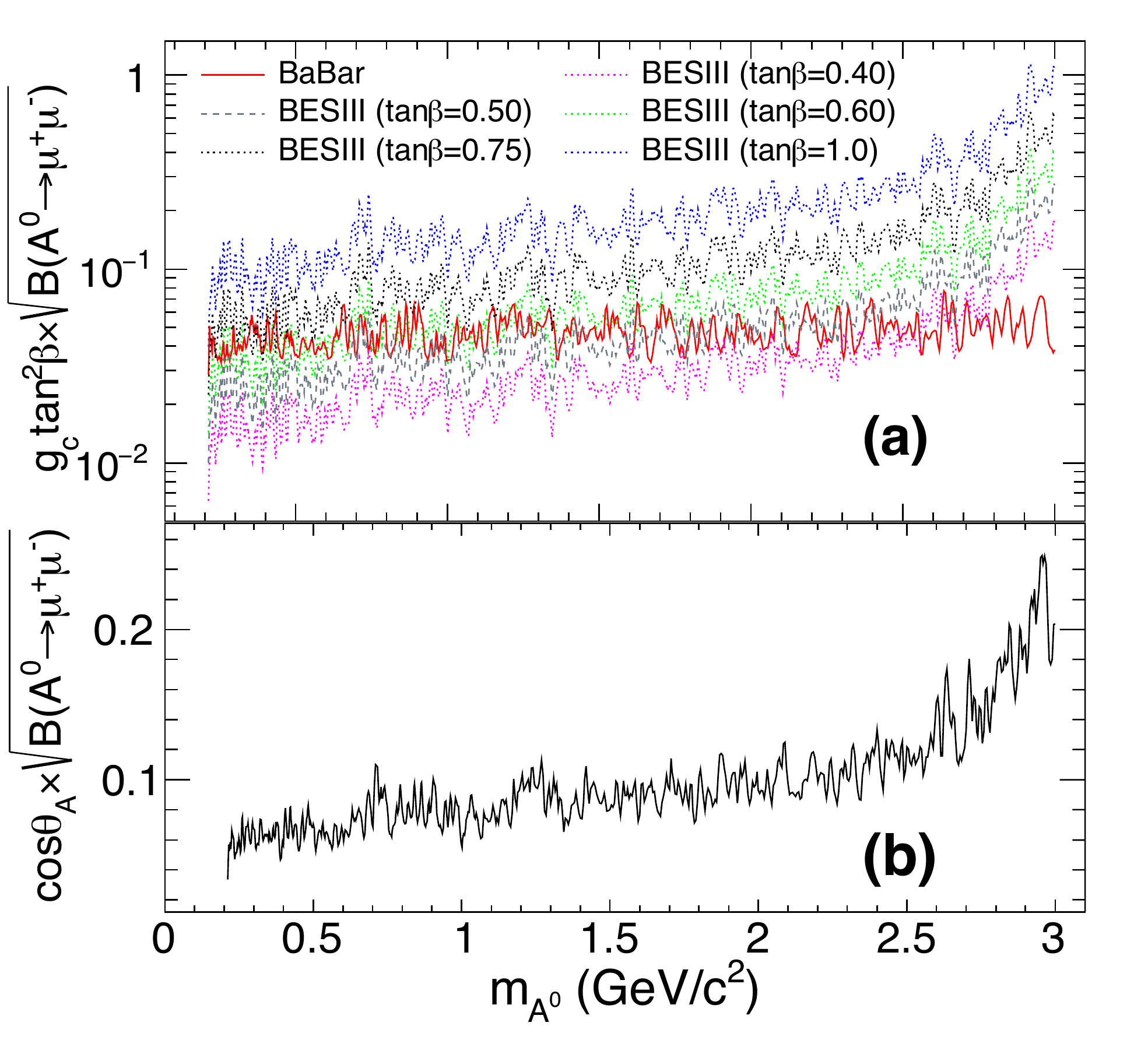}
\caption{The $90\%$ C.L. upper limits on product branching fractions $\mathcal{B}(J/\psi \to \gamma A^0)\times \mathcal{B}(A^0 \to \mu^+\mu^-)$ (left), $g_b(=g_c\tan^2\beta)\times \sqrt{\mathcal{B}(A^0 \to \mu^+\mu^-)})$ for the BaBar~\cite{babarhiggs} and BESIII~\cite{bes3higgs} measurements (right top) and $\cos\theta_A(=|\sqrt{g_bg_c}|\times \sqrt{\mathcal{B}(A^0 \to \mu^+\mu^-)})$ (right bottom) as a function of $m_{A^0}$ including all the systematic uncertainties. In left hand side figure, the inner and outer bands include the statistical uncertainty only and contain $68\%$ and $95\%$ of the expected limits values that are computed using a large number of pseudoexperiments.}
\label{mred}
\end{figure}

The search for $A^0$ is also performed using 225 million $J/\psi$ events collected at the $J/\psi$ resonance by the BESIII detector~\cite{bes3higgs}. The background is mainly dominated by the non-resonant component of $e^+e^- \to \gamma \mu^+\mu^-$ process and the resonant components of $J/\psi \to \rho \pi$ and $J/\psi \to \gamma f$ ($f=f_2(1270), f_0(1710)$) decays. We perform the search for a narrow resonance by performing the ML fit to the reduced mass distribution, defined $m_{\rm red} = \sqrt{m_{\mu^+\mu^-}^2-4m_{\mu}^2}$, of data at 2035 mass points in the steps of 1-2 MeV/$c^2$, where $m_{\mu^+\mu^-}$ is the di-muon invariant mass distribution and $m_{\mu}$ is the nominal muon mass~\cite{pdg}. No evidence of $A^0$ production is found at any $m_{A^0}$ point.  We set the $90\%$ C.L. upper limits on product branching fractions $\mathcal{B}(J/\psi \to \gamma A^0)\times \mathcal{B}(A^0 \to \mu^+\mu^-)$  that vary in the range of  $(2.8-495.3)\times 10^{-8}$ for $0.212 \le m_{A^0} \le 3.0$ GeV/$c^2$ (Figure~\ref{mred} (left)). This new result has five times improvement over the previous BESIII measurement~\cite{A0psip}. The upper limit on  $g_b(=g_c\tan^2\beta)\times \sqrt{\mathcal{B}(A^0 \rightarrow \mu^+\mu^-)}$ is also computed for different values of $\tan\beta$ at $90\%$ C.L. in order to compare this result with the BaBar measurement~\cite{babarhiggs}. Our result is  better than the BaBar measurement~\cite{babarhiggs} in the low-mass region for $\tan \beta \le 0.6$ (Figure~\ref{mred} (right)). The combined results of both BaBar and BESIII in term of $\cos\theta_A \times \sqrt{\mathcal{B}(A^0 \to \mu^+\mu^-)}$, which is independent of $\tan \beta$, reveal that the nature of the $A^0$ is mostly singlet in nature. 

\section{Summary}
BESIII has performed the searches for di-lepton decays of light Higgs and dark bosons as well as invisible decays of light vector and pseudoscalar mesons  using the data samples collected at $J/\psi$, $\psi(3686)$ and $\psi(3770)$ resonances. No evidence of any significant signal events is found in these data-sets and set one of the stringent exclusion limits. These exclusion limits may constrain a large fraction of the parameters of the new physics models including the NMSSM and hidden dark-sector.  
  BESIII is performing the searches for new physics in several other decay modes too using the large data-sets collected at various energy points, including  10 billion $J/\psi$ events that have recently collected by the BESIII detector, and we look forward to seeing more exciting results on the new physics topics in the near future.

\end{document}